\documentclass[11pt,onecolumn]{IEEEtran}

\usepackage[utf8]{inputenc} 
\usepackage[T1]{fontenc}
\usepackage{url}
\usepackage{ifthen}
\usepackage{cite}
\usepackage[cmex10]{amsmath} % Use the [cmex10] option to ensure complicance
\usepackage{amsfonts}
\usepackage{amssymb}
\usepackage{amsthm}
\usepackage{algorithmicx}
\usepackage{algpseudocode}
\usepackage{algorithm}
\usepackage{subfigure}
\usepackage{graphicx}
\usepackage{cite}
\usepackage{calc}
\usepackage{color}
\usepackage{epsfig}
\usepackage{setspace}
\usepackage{multirow}
\usepackage{mathtools, cuted}
\usepackage{epstopdf}
\usepackage{lipsum}
\usepackage{mathtools}
\usepackage{cuted}
\usepackage{caption}

\newtheorem{defn}{Definition}
\newtheorem{thm}{{\cal T}heorem}
\newtheorem{cor}{Corollary}
\newtheorem{prop}{Proposition}
\newtheorem{lem}{Lemma}
\newtheorem{conj}{Conjecture}
\newtheorem{constr}{Construction}
\newtheorem{note}{Remark}
\newcommand{\bit}{\begin{itemize}}
	\newcommand{\eit}{\end{itemize}}
\newcommand{\bcor}{\begin{cor}}
	\newcommand{\ecor}{\end{cor}}
\newcommand{\beq}{\begin{equation}}
\newcommand{\eeq}{\end{equation}}
\newcommand{\beqn}{\begin{equation}}
\newcommand{\eeqn}{\end{equation}}
\newcommand{\bea}{\begin{eqnarray}}
\newcommand{\eea}{\end{eqnarray}}
\newcommand{\bean}{\begin{eqnarray*}}
	\newcommand{\eean}{\end{eqnarray*}}
\newcommand{\ben}{\begin{enumerate}}
	\newcommand{\een}{\end{enumerate}}
\newcommand{\bdefn}{\begin{defn}}
	\newcommand{\edefn}{\end{defn}}
\newcommand{\bnote}{\begin{note}}
	\newcommand{\enote}{\end{note}}
\newcommand{\bprop}{\begin{prop}}
	\newcommand{\eprop}{\end{prop}}
\newcommand{\blem}{\begin{lem}}
	\newcommand{\elem}{\end{lem}}
\newcommand{\bthm}{\begin{thm}}
	\newcommand{\ethm}{\end{thm}}
\newcommand{\bconj}{\begin{conj}}
	\newcommand{\econj}{\end{conj}}
\newcommand{\bconstr}{\begin{constr}}
	\newcommand{\econstr}{\end{constr}}
\newcommand{\bpf}{\begin{proof}}
	\newcommand{\epf}{\end{proof}}
\newcommand{\bprf}{{\em Proof: }}
\newcommand{\eprf}{\hfill $\Box$}

\newcommand{\params}{\mbox{$(a,b,\tau)$}}

\IEEEoverridecommandlockouts 
\begin{document}
	
\title{Explicit Rate-Optimal Streaming Codes \\ with Smaller Field Size} 
\author{%
	\IEEEauthorblockN{	Myna Vajha\IEEEauthorrefmark{1},
		Vinayak Ramkumar\IEEEauthorrefmark{1},
	    M. Nikhil Krishnan\IEEEauthorrefmark{2}, P. Vijay Kumar\IEEEauthorrefmark{1}\\}
	\IEEEauthorblockA{\IEEEauthorrefmark{1}%
		Department of Electrical Communication Engineering, IISc Bangalore\\}
	\IEEEauthorblockA{\IEEEauthorrefmark{2}%
		Department of Electrical and Computer Engineering, University of Toronto\\}
	\IEEEauthorblockA{
		\{mynaramana, vinram93, nikhilkrishnan.m, pvk1729\}@gmail.com}
\thanks{%P. Vijay Kumar is also a  Visiting  Professor  at  the  University  of  Southern  California.
This  research  is  supported by  the J C Bose National Fellowship JCB/2017/000017.}
}	

	\maketitle
	
\begin{abstract}
Streaming codes are a class of packet-level erasure codes that ensure packet recovery over a sliding window channel which allows either a burst erasure of size $b$ or $a$ random erasures within any window of size $(\tau+1)$ time units, under a strict decoding-delay constraint $\tau$. The field size over which streaming codes are constructed is an important factor determining the complexity of implementation. The best known explicit rate-optimal streaming code requires a field size of $q^2$ where $q \ge \tau+b-a$ is a prime power. In this work, we present an explicit rate-optimal streaming code, for all possible $\{a,b,\tau\}$ parameters, over a field of size $q^2$ for prime power $q \ge \tau$.  This is the smallest-known field size of a general explicit rate-optimal construction that covers all  $\{a,b,\tau\}$ parameter sets.  We achieve this by modifying the non-explicit code construction due to Krishnan et al. to make it explicit, without change in field size.         
\end{abstract}

%% The paper must be self-contained. However, if you are referring to
%% a full version for checking certain proofs, please provide the
%% publically accessible location below.  If the paper is completely
%% self-contained, you can remove the following line from your
%% submission.
%\textit{A full version of this paper is accessible at \cite{Small}.}
%\url{https://arxiv.org/pdf/21xx.xxxx.pdf} 

\section{Introduction\label{sec:intro}}	%-------------------------------------------------------------------------------

%\section{Background\label{sec:background}}
Enabling low-latency reliable communication for applications such as telesurgery, industrial automation, augmented reality and vehicular communication is a key target of 5G communication systems. For instance, telesurgery camera flow requires packet-loss rate less than $10^{-3}$ and end-to-end latency below $150$ ms  \cite{5GAmericas}. To combat the packet drops that are an inevitable part of any communication network, one approach is to employ feedback-based methods such as ARQ. But such feedback-based schemes incur round-trip propagation delay, making it difficult to meet low-latency requirements. Blind re-transmission of packets is another option, but is inefficient as it amounts to using a repetition code. Streaming codes are a class of packet-level erasure codes and represent a natural way of achieving reliable, low-latency communication at the packet level.  

A packet-expansion encoding framework for streaming codes was introduced in \cite{MartSunTIT04}. Given message packet at time $t$ denoted by $\underline{u}(t) \in \mathbb{F}_q^k$, the coded packet $\underline{x}(t) \in \mathbb{F}_q^n$, at time $t$, is generated by appending parity packet $\underline{p}(t)\in \mathbb{F}_q^{n-k}$ to $\underline{x}(t)$. More formally, $\underline{x}(t) = \left[\underline{u}(t)^T \ \underline{p}(t)^T\right]^T$. The encoder is causal and hence $\underline{p}(t)$ depends only on $\underline{u}(t)$ and prior message packets. In \cite{MartSunTIT04,MartTrotISIT07}, streaming codes that can handle burst erasure of size $b$ under a  decoding-delay constraint $\tau$ are presented. The decoding-delay constraint $\tau$ means that for recovery of message packet $\underline{u}(t)$ only packets with index $ \le t+\tau$ can be accessed. In \cite{BadrPatilKhistiTIT17}, Badr et al. presented a delay constrained sliding window (DCSW) channel model that allows burst or random erasures. This channel can be viewed as a deterministic approximation of the Gilbert-Elliot channel \cite{VajRamJhaKum}.
%, which is a widely used channel model for network erasures \cite{HasHoh,HohGeiHas}. 
In the DCSW channel model, within any sliding window of size $w$ time units, either a burst erasure of length $\leq b$ or else, at most $a$ random erasures can occur. Additionally there is a decoding-delay constraint $\tau$. This model is non-trivial only if $0<a \leq b \leq \tau$. As it turns out, we can, without loss in generality, set $w=\tau+1$ (see \cite{NikDeepPVK}).  Thus the DCSW channel is parameterized by the three-parameter set $\{a,b,\tau\}$. An \params\ streaming code is a packet-level code that can recover from all the permissible erasure patterns of an $\{a,b,\tau\}$ DCSW channel, within decoding-delay $\tau$. 
Some other models of erasure codes for streaming can be found in \cite{AdlCas,LeongHo,LeoQurHo,TekHoYaoJag}.

In \cite{BadrPatilKhistiTIT17} an upper bound on the rate of an \params\ streaming code was presented and it  was later shown in  \cite{FongKhistiTIT19, NikPVK} that this rate is achievable for all possible $\{a,b,\tau\}$ parameters. It follows from these results that the optimal rate of \params\ streaming code is given by $R_{\text{opt}} = \frac{\tau+1-a}{\tau+1-a+b}$. The rate-optimal codes presented in \cite{FongKhistiTIT19, NikPVK} required a finite field alphabet that is exponential in $\tau$. A non-explicit rate-optimal streaming code, which requires a field $\mathbb{F}_{q^2}$ with prime power $q \ge \tau$, is presented in \cite{NikDeepPVK}. Subsequently,  an explicit construction was presented in \cite{KhistiExplicitCode} requiring field size $q^2$, for $q \ge \tau+b-a$, a prime power. Streaming codes for variable packet sizes are explored in \cite{RudRas}. Explicit rate-optimal constructions having linear field size for some $\{a,b,\tau\}$ parameter ranges are presented in \cite{NikDeepPVK, NikRamVajKum, RamVajNikPVK}. However, the construction in \cite{KhistiExplicitCode} remains the smallest field size explicit rate-optimal streaming code construction that exists for all possible $\{a,b,\tau\}$. Note that the field size required for the explicit code construction in \cite{KhistiExplicitCode} is larger than the field size $q^2 \geq \tau^2$ requirement of the code in \cite{NikDeepPVK}. In the present paper, we present an explicit rate-optimal code having the same field-size requirement $q^2$, with prime power $q \ge \tau$, as that of the non-explicit code in \cite{NikDeepPVK}. Smaller field size constructions simplify implementation and are hence of significant, practical interest. 

The principal contribution of the paper is thus an explicit rate-optimal streaming code construction for all possible $\{a,b,\tau\}$ parameters.  The construction is motivated by the structure of the non-explicit code in \cite{NikDeepPVK} and has smallest known field size of an explicit rate-optimal streaming code construction that holds for all $\{a,b,\tau\}$ parameters.  

Section~\ref{Sec:Pre} presents the diagonal-embedding framework for embedding a scalar code within the packet stream. The explicit construction of the scalar code having field size $q^2 \geq \tau^2$ is presented in Section~\ref{Sec:Const}.  Proof that this construction, in conjunction with diagonal embedding, results in a rate-optimal streaming code is presented in Section~\ref{Sec:Proof}.  

We use $[a:b]$ to denote the set $\{a,a+1, \dots, b-1, b\}$. 
Given a $(k \times n)$ matrix $M$, $I \subseteq [0:k-1]$ and $J \subseteq [0:n-1]$,  $M(I,J)$ will denote the sub-matrix of $M$ comprised of rows with row-index in $I$ and columns with column-index in $J$. We use $|M|$ to denote the determinant of $M$, $I_{u}$ denotes $(u \times u)$ identity matrix and  $\underbrace{\mathbf{0}}_{(u \times v)}$ will denote the $(u \times v)$ all-zero matrix.

\section{Preliminaries} \label{Sec:Pre}
\subsection{Diagonal Embedding}
\begin{figure}[ht!]
	\captionsetup{font=footnotesize}
	\begin{center}
		\includegraphics[width=2.7in]{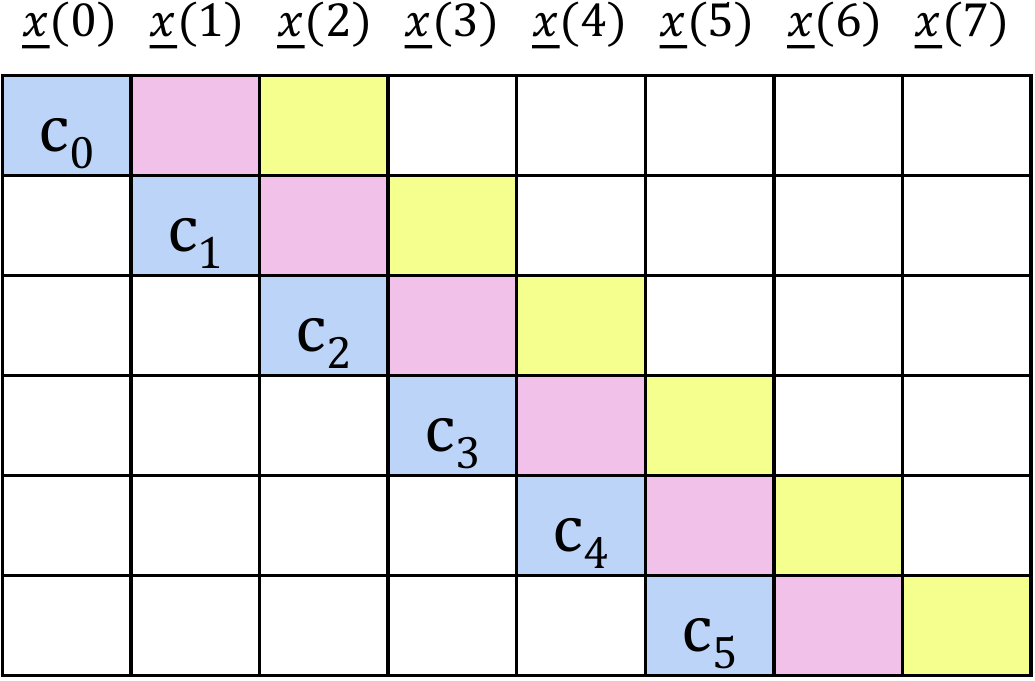}
		\caption{Packet-level code constructed by diagonal embedding of a scalar code of block length $6$. Here each column indicates a coded packet.}
		\label{fig:diag_example}
	\end{center}
\vspace{-0.5cm}
\end{figure}

Diagonal embedding, introduced in \cite{MartTrotISIT07}, can be viewed as a framework for deriving a packet-level code from a scalar code.  This technique has been consistently used in the streaming-code literature.  Let $\mathcal{C}$ be an $[n,k]$ scalar code in systematic form, with first $k$ code symbols being message symbols. Consider a packet-level code with coded packet at time $t$ denoted by $\underline{x}(t)=[x_0(t)~x_1(t)\dots x_{n-1}(t)]^T$.  We will say that the packet-level code is obtained by diagonal embedding of the scalar code $\mathcal{C}$ if for all $t$, each $n$-tuple $\left(x_0(t),x_1(t+1), \dots,x_{n-1}(t+n-1)\right)$ is a codeword in the scalar code $\mathcal{C}$.  The packet-level code shares the rate $\frac{k}{n}$ of the underlying scalar code.   Diagonal embedding is illustrated in Fig.\ref{fig:diag_example}. 
\subsection{Properties Required of the Scalar Code}

Let $\delta=b-a$. In order to show that the packet-level code constructed through diagonal embedding of an $[n=\tau+1+\delta,k=n-b]$ scalar code $\mathcal{C}$ is a rate-optimal $(a, b, \tau)$ streaming code it suffices to show that the following erasure recovery properties hold for codeword  $(c_0,c_1,\dots, c_{n-1}) \in \mathcal{C}$  that occupies the time indices $0 \leq t \leq (n-1)$.  Analogous, time-shifted versions of these below conditions apply to the other embedded codewords (see \cite{NikDeepPVK} for details). 
\ben
\item[B1] Any code symbol $c_t$ with $t \in [0:\delta-1]$ should be recoverable from the erasure of a burst of $b$ packets, corresponding to time indices in $[t:t+b-1]$, by accessing non-erased code symbols in the set $\{c_i \mid t < i \le \tau+t \} \cup \{c_i \mid i < t \}$.  The latter set represents previously-decoded code symbols. 
\item[R1] Any code symbol $c_t$ with $t \in [0:\delta-1]$ should be recoverable from any $a$ random packet erasures, corresponding to time indices $t$ and $(a-1)$ indices in $[t+1:\tau+t]$, by accessing non-erased code symbols in the set $\{c_i \mid t< i \le \tau+t \} \cup \{c_i \mid i < t \}$.
\item[B2] For any $t \in [\delta:\tau+1-a]$, code symbols $\{ c_i \mid i \in [t:t+b-1]\}$ should be recoverable by accessing remaining code symbols $\{c_i \mid i \notin [t:t+b-1]\}$.
%\item[R2] For any $t \in [\delta:\tau+\delta]$ and set $A \subseteq [t:n-1]$ with $t= \min(A)$, $|A|=a$, code symbols $\{ c_i \mid i \in A \}$ should be recoverable by accessing the remaining code symbols $\{c_i \mid i \notin A\}$.
\item[R2] For any set $A \subseteq [\delta:\tau+\delta]$ with $|A|=a$, code symbols $\{ c_i \mid i \in A \}$ should be recoverable by accessing the remaining code symbols $\{c_i \mid i \notin A\}$.
\een
%The property B2 guarantees systematic encoding since $n-k=b$ parity symbols $\{c_k,\dots,c_{n-1}\}$ can be computed from $k$ message symbols $\{c_0,\dots, c_{k-1}\}$.
It follows from B2 property for $t=\tau+1-a$ that the last $n-k=b$ code symbols $\{c_k,\dots,c_{n-1}\}$ can be computed from the first $k$ code symbols $\{c_0,\dots, c_{k-1}\}$, thereby  guaranteeing systematic encoding with $\{c_0,\dots, c_{k-1}\}$ as message symbols.

\section{Scalar Code Construction} \label{Sec:Const}

Our explicit, rate-optimal streaming code construction will employ diagonal embedding as well as an $[n=\tau+1+\delta,k=n-b]$ scalar code satisfying the four erasure recovery properties listed above in Section~\ref{Sec:Pre}. We  begin by recursively defining a matrix that will be used to specify the parity-check matrix of our scalar code. This recursive definition can be viewed as an extension of the recursive matrix definition in \cite{HolTol} that was used to construct rate-optimal binary streaming codes for the situation when only burst erasures are present. 
%In \cite{HolTol}, explicit binary rate-optimal streaming codes, dealing only with burst erasures were obtained using that matrix.  
\bdefn \label{def:Pmatrix}
For any positive integers $u, v$ and $a$, we recursively define the $(u \times v)$ matrix $\mathbf{P}_{u, v}^a$ as shown below:
\bean
\mathbf{P}_{u, v}^{a} &=& \begin{cases}
	\left[ \begin{array}{ccc}
		I_{u} & \underbrace{\mathbf{0}}_{(u \times a)} & \mathbf{P}_{u, v-u-a}^a
	\end{array} \right] &  u+a < v\\ \ \\
	\left[ \begin{array}{cc}
		I_{u} & \underbrace{\mathbf{0}}_{(u \times (v-u))} 
	\end{array} \right] & u \le v \le u+a\\ \ \\
	%	\left[ \begin{array}{c}
	%		I_{u} \end{array} \right] & u = v\\ \ \\
	\left[ \begin{array}{c}
		I_{v} \\
		\mathbf{P}_{u-v, v}^a 
	\end{array} \right] & v < u\\
\end{cases}
\eean
\edefn
For example, 
$\mathbf{P}^2_{3,7}=	[I_{3}~\underbrace{\mathbf{0}}_{(3 \times 2)}~\mathbf{P}_{3, 2}^2],~\mathbf{P}_{3, 2}^2= \left[ \begin{array}{c}
I_{2} \\
\mathbf{P}_{1, 2}^2 
\end{array} \right] ~\text{and}~ \mathbf{P}_{1, 2}^2 =[1~0]$.
Therefore we have 
\bean 
\mathbf{P}^2_{3,7}= \left[ \begin{array}{ccc|cc|cc}
	1 & 0 & 0 & 0 & 0 & 1 & 0   \\
	0 & 1 & 0 & 0 & 0 & 0 & 1  \\ \cline{6-7}
	0 & 0 & 1 & 0 & 0 & 1 & 0  
\end{array}\right].
\eean 
\begin{constr} \label{constr}\normalfont
	Let $\delta = b-a$, $q \ge \tau$ be a prime power and $\alpha \in \mathbb{F}_{q^2} \setminus \mathbb{F}_q$. Let the $(a \times (\tau+1-a))$ matrix $\mathbf{C}$ over $\mathbb{F}_q$ be such that any square sub-matrix of it is non-singular. We define an $[n=\tau+1+\delta, k = n-b]$ scalar code having a parity check matrix $H$ that is defined in step-by-step fashion below:       
	\bit
	\item initialize $H$ to be the $(b \times (\tau+1+\delta))$ all-zero matrix, 
	\item set $H([0:\delta-1],[0:\delta-1]) \ = \ \alpha I_{\delta}$,
	\item set $H([0:\delta-1],[b:\tau-1]) \ = \ \mathbf{P}^a_{\delta,\tau-b}$,
	\item set $H([\delta:b-1],[0:a-1]) \ = \ I_{a}$,
	\item set $H([\delta:b-1],[a:\tau]) \ = \ \mathbf{C}$,
	\item set $H(0,\tau)=\alpha$ and $H([1:\delta],[\tau+1:\tau+\delta])=I_{\delta}$.
	\eit 	  
\end{constr}

The first $\delta$ rows of the parity check matrix $H$ are given by:
\bean
H([0:\delta - 1], [0:\tau+\delta]) 
\eean 
\bean 
= \left[ \begin{array}{cccc|c|c|ccccc}
	\alpha &  & & & \multirow{4}{*}{$\underbrace{\mathbf{0}}_{(\delta \times a)}$} & \multirow{4}{*}{$\underbrace{\mathbf{P}_{\delta, \tau-b}^a}_{(\delta \times (\tau-b))}$} & \alpha & & & & 0\\
	& \alpha & & & & & & 1 & & & 0\\
	& & \ddots & & & & & & \ddots & & \vdots\\
	& &  & \alpha & & & & & & 1 & 0
	\end{array}\right],
\eean
and the last $(b-\delta)=a$ rows of $H$ by: 
\bean
H([\delta:b - 1], [0:\tau+\delta])  
\eean 
\bean 
= \left[ \begin{array}{cccc|c|c|c}
	1 &  & & & \multirow{4}{*}{$\underbrace{\mathbf{C}}_{(a \times (\tau+1-a))}$} & \multirow{4}{*}{$\underbrace{\mathbf{0}}_{(a \times (\delta-1) )}$} & 1\\
	& 1 & & & & & 0\\
	& & \ddots & & & & \vdots\\
	& &  & 1 & & & 0
\end{array}\right].
\eean
As defined above, $H([\delta:b-1],[0:\tau])$ is the parity check matrix of a $[\tau+1,\tau+1-a]$ MDS code.  A finite field of size $q \ge \tau$ suffices to explicitly construct the matrix $\mathbf{C}$.  It can be verified that the last $a$ rows of $H$ are the same as that of the non-explicit code presented in \cite{NikDeepPVK}.

\subsection{Example Constructions}
\subsubsection{$(a=2,b=5,\tau=12)$} 
Here $\delta=3$, $\tau-b=7$, $\tau+1-a=11$ and $\tau+1+\delta=16$. The parity check matrix $H$ of $[n=16,k=11]$ scalar code is given in this case by:
\bean
	\left[ \begin{array}{ccc|cc|ccccccc|cccc}
	\alpha & 0 & 0 & 0 & 0 & 1 & 0 & 0 & 0 & 0 & 1 & 0 & \alpha & 0 & 0 & 0  \\
	0 & \alpha & 0 & 0 & 0 & 0 & 1 & 0 & 0 & 0 & 0 & 1 & 0 & 1 & 0 & 0  \\
	0 & 0 & \alpha & 0 & 0 & 0 & 0 & 1 & 0 & 0 & 1 & 0 & 0 & 0 & 1 & 0  \\
	\hline 
	1 & 0  &  \multicolumn{11}{|c|}{ \multirow{2}{*}{$\underbrace{\mathbf{C}}_{(2 \times 11)}$}} & 0 & 0 & 1 \\
	0 & 1  & \multicolumn{11}{|c|}{} & 0 & 0 & 0
	\end{array}\right].
\eean
\subsubsection{$(a=3,b=6,\tau=8)$} Here $\delta=3$, $\tau-b=2$, $\tau+1-a=6$ and $\tau+1+\delta=12$. The parity check matrix $H$ of $[n=12,k=6]$ scalar code is given in this case by:
\bean
\left[ \begin{array}{ccc|ccc|cc|cccc}
	\alpha & 0 & 0 & 0 & 0 & 0 & 1 & 0 & \alpha & 0 & 0 & 0  \\
	0 & \alpha & 0 & 0 & 0 & 0 &  0 & 1 & 0 & 1 & 0 & 0  \\
	0 & 0 & \alpha & 0 & 0  & 0 & 1 & 0 & 0 & 0 & 1 & 0  \\
	\hline 
	1 & 0  & 0 & \multicolumn{6}{c|}{ \multirow{3}{*}{$\underbrace{\mathbf{C}}_{(3 \times 6)}$}} & 0 & 0 & 1 \\
	0 & 1  & 0 & \multicolumn{6}{|c|}{} & 0 & 0 & 0 \\
	0 & 0  & 1 & \multicolumn{6}{|c|}{} & 0 & 0 & 0
\end{array}\right]. \\
\eean

\section{Proof of Erasure Recovery Properties} \label{Sec:Proof}
In this section we show that the $[n=\tau+1+\delta,k=n-b]$ scalar code defined in Section~\ref{Sec:Const} satisfies all the four erasure recovery conditions. This will in turn prove that the packet-level code obtained by diagonal embedding of this scalar code is a rate-optimal $(a,b,\tau)$ streaming code.  This $(a,b,\tau)$ streaming code can be  explicitly constructed over a finite field of size $q^2\geq \tau^2$.  

%It can be proved that the scalar code satisfies R1 and R2 properties using arguments similar to that presented in~\cite{NikDeepPVK}. Proof of the R1 and R2 properties, can be found in the complete version of this paper at \cite{Small}. Proof of properties B1 and B2 appear below. 

It can be proved that the scalar code satisfies R1 and R2 properties using arguments similar to that presented in~\cite{NikDeepPVK}. Nevertheless for the sake of completeness, we provide proof of all the four properties here.

\subsection{Proof of B1 Property}

Property B1 is verified by showing that for every $t \in [0:\delta-1]$, there exists a parity check equation having  support at $t$ and zeros at indices $[t+1:t+b-1] \cup [t+\tau+1:\tau+\delta]$.  Using this parity-check equation, code symbol $c_t$ can be recovered from a burst erasure confined to $[t:t+b-1]$ by accessing only the non-erased code symbols having index $\le t+\tau$. %We will show the existence of such a parity check equation case by case. 

For the $(a=2,b=5,\tau=12)$ example, suppose symbols $\{c_0,c_1,c_2,c_3,c_4\}$ are erased. Row $0$ of the parity check matrix $H$ for this example is as shown below:
\bean
\left( \begin{array}{cccccccccccccccc}
	\alpha & 0 & 0 & 0 & 0 & 1 & 0 & 0 & 0 & 0 & 1 & 0 & \alpha & 0 & 0 & 0  
	\end{array}\right).
\eean
It follows that $\alpha c_{0}+c_5+c_{10}+\alpha c_{12}=0$. Hence $c_0$ can be recovered by accessing symbols till $c_{12}$. Now for the $(a=3,b=6,\tau=8)$ example, consider a burst erasure such that $\{c_2,c_3,c_4,c_5,c_6,c_7\}$ are lost. It follows from row $2$ of $H$ that $\alpha c_{2}+c_{6}+c_{10}=0$, but since $c_6$ is erased this equation alone is not sufficient to recover $c_2$. We get $\alpha c_{0}+c_{6}+\alpha c_{8}=0$ from the $0$-th row of $H$. From these two parity-check equations we obtain $\alpha c_{2}+c_{10}-\alpha c_{0}-\alpha c_{8}=0$, using which $c_2$ can be recovered by accessing code symbols till $c_{10}$. We now prove the B1 property for general $\{a,b,\tau\}$. We use the notation $h(t)$ to denote the $t$-th row of $H$. The symbol 'X' is used as a don't care symbol in the arguments below. 

\subsubsection{$\tau-b \ge \delta$} From the definition of the matrix $P_{\delta, \tau-b}^a$, the first $\delta$ columns of $P_{\delta, \tau-b}^a$ form $I_{\delta}$. The $t$-th row of $H$ looks as shown below:
\begin{equation} \nonumber 
\begin{array}{ccccccccccccccccccc}
	0 &  & &  t & & & & t+b & & & &  b+\delta &  & \tau+t & & &  \tau+\delta \\
	(0 &\cdots& 0 & \alpha & 0 & \cdots & 0 & 1 & 0 & \cdots & 0 & X & \cdots & X & 0 & \cdots & 0)
	\end{array}
\end{equation}
Therefore, using the parity check equation given by $h(t)$ we can recover code symbol $c_t$ from a burst erasures at $[t:t+b-1]$ by accessing available code symbols with index $\le \tau+t$. 
%\bean
%\begin{array}{ccccccccccccccccccc}
%	& 0 & \multicolumn{2}{c}{\cdots} &  t &   \multicolumn{3}{c}{\cdots} & (t+b) & \multicolumn{3}{c}{\cdots} & (b+\delta) & \cdots & (\tau+t) & \multicolumn{2}{c}{\cdots} & (\tau+\delta) &\\
%	h(t)=(& 0 & \cdots & 0 & \alpha & 0 & \cdots & 0 & 1 & 0 & \cdots & 0 & X & \cdots & X & 0 & \cdots & 0 & )
%\end{array}
%\eean
%Therefore it is clear to see that by using the $t$-th parity check equation code symbol $c_t$ can be recovered from a burst erasures at $[t:t+b-1]$ by accessing code symbols with index $\le \tau+t$.
\subsubsection{$\tau-b < \delta$} Let $\ell=\tau-b$, $\delta = v \ell + x$ where $0 \le x < \ell$. Then by definition $P_{\delta, \ell}^a = \left[I_{\ell}~\cdots~I_{\ell}~{P_{x, \ell}^a}^T
\right]^T.$
%\bean
%P_{\delta, \ell}^a = \left[ \begin{array}{c}
%	I_{\ell}\\
%	%I_{\ell}\\
%	\vdots\\
%	I_{\ell}\\
%	P_{x, \ell}^a
%\end{array}
%\right].
%\eean
We further divide the proof of this case into three sub-cases. 
%We will now show that for any $t \in [0:\delta-1]$ there exists a parity check equation using which $c_t$ can be recovered from a burst of size $b$ starting at $t$.

\paragraph{$t \in [0:\ell-1]$} In this case, the $t$-th row of $H$ satisfies the requirement as shown below and hence can be used for recovery of $c_t$.
\begin{equation} \nonumber 
\begin{array}{cccccccccccccccc}
	0 &  & &  t & & & & t+b & & & &  \tau+t &  & &   \tau+\delta \\
	(0 &\cdots& 0 & \alpha & 0 & \cdots & 0 & 1 & 0 & \cdots & 0 & X & 0 & \cdots & 0)
	\end{array}
\end{equation}
%\bean
%\begin{array}{ccccccccccccccccccc}
%	& 0 & \multicolumn{2}{c}{\cdots} &  t &   \multicolumn{3}{c}{\cdots} & (t+b) & \multicolumn{3}{c}{\cdots} & (\tau+t) & \multicolumn{2}{c}{\cdots} & (\tau+\delta) &\\
%	h(t)=(& 0 & \cdots & 0 & \alpha & 0 & \cdots & 0 & 1 & 0 & \cdots & 0 & X & 0 & \cdots & 0 & )
%\end{array}
%\eean
%Therefore $c_t$ can be recovered from the burst of size $b$ starting at $t$ by accessing code symbols with index $\le \tau+t$.
\paragraph{$t \in [\ell:v\ell-1]$} Let $t = v' \ell + x'$ where $1 \le v' \le v-1$ and $0 \le x' < \ell$. The $t$-th row  $h(t)$ is of the form:
\begin{equation} \nonumber 
\begin{array}{ccccccccccccccccc}
	0 &  & &  t & & & & b+x' & & & &  \tau+t &  & & \tau+\delta \\
	(0 &\cdots& 0 & \alpha & 0 & \cdots & 0 & 1 & 0 & \cdots & 0 & X & 0 & \cdots & 0)
	\end{array}
\end{equation}
This parity check equation does not have $(b-1)$ zeros following the index $t$. Therefore we look at the equation given by $(t-\ell)$-th row of $H$:
\begin{equation} \nonumber 
\begin{array}{ccccccccccccccccc}
	0 &  & &  t-\ell & & & & b+x' & & & &  \tau+t-\ell &  & & \tau+\delta \\
	(0 &\cdots& 0 & \alpha & 0 & \cdots & 0 & 1 & 0 & \cdots & 0 & X & 0 & \cdots & 0)
	\end{array}
\end{equation}
%\bean
%\begin{array}{ccccccccccccccccc}
%	& 0 & \multicolumn{2}{c}{\cdots} &  (t-\ell) &   \multicolumn{3}{c}{\cdots} & (b+x_1) & \multicolumn{3}{c}{\cdots} &  (\tau+t-\ell) & \multicolumn{2}{c}{\cdots} & (\tau+\delta) &\\
%	h(t-\ell) = ( & 0 & \cdots & 0 & \alpha & 0 & \cdots & 0 & 1 & 0 & \cdots & 0 & X & 0 & \cdots & 0 &)
%\end{array}
%\eean
Thus we get a parity check $h(t)-h(t-\ell)$ as shown below:
\begin{equation} \nonumber 
\begin{array}{ccccccccccccccccc}
	0 &  & &  t & & & & \tau+t-\ell & & & &  \tau+t &  & & \tau+\delta \\
	(X &\cdots& X & \alpha & 0 & \cdots & 0 & X & 0 & \cdots & 0 & X & 0 & \cdots & 0)
	\end{array}
\end{equation}
%\bean
%\begin{array}{ccccccccccccccccc}
%	& 0 & \multicolumn{2}{c}{\cdots} &  t &   \multicolumn{3}{c}{\cdots} & (\tau+t-\ell) & \multicolumn{3}{c}{\cdots} &  (\tau+t) & \multicolumn{2}{c}{\cdots} & (\tau+\delta) &\\
%	( & X & \cdots & X & \alpha & 0 & \cdots & 0 & 1 & 0 & \cdots & 0 & X & 0 & \cdots & 0 &)
%\end{array}
%\eean
Note that $\tau+t-\ell = (\tau+t)-(\tau-b) = t+b$. Therefore there are $(b-1)$ zeros following index $t$ in the parity check equation shown above and this parity check equation can be used to recover code symbol $c_t$.
\paragraph{$t \in [v\ell:v\ell+x-1]$} Let $t=v\ell+x'$, where $x' \le x-1$. The $t$-th parity check equation $h(t)$ is of the form:
\begin{equation} \nonumber 
\begin{array}{cccccccccccccccccccc}
	0 &  & &  t & & & & b+x' & & & &  \tau &  & & \tau+t & & & \tau+\delta \\
	(0 &\cdots& 0 & \alpha & 0 & \cdots & 0 & 1 & X & \cdots & X & 0 & \cdots  & 0 & X & 0 & \cdots & 0)
	\end{array}
\end{equation}
%\bean
%\begin{array}{cccccccccccccccccccc}
%	& 0 & \multicolumn{2}{c}{\cdots} &  t &   \multicolumn{3}{c}{\cdots} & (b+x_1) & \multicolumn{3}{c}{\cdots} &  \tau & \multicolumn{2}{c}{\cdots} & (\tau+t) & \multicolumn{2}{c}{\cdots} & (\tau+\delta) &\\
%	h(t)=( & 0 & \cdots & 0 & \alpha & 0 & \cdots & 0 & 1 & X & \cdots  & X & 0 & \cdots & 0 & X & 0 & \cdots & 0 &)
%\end{array}
%\eean
as the first $x$ columns of $P_{x, \ell}^a$ are given by $I_x$. Let $y_i= (v-1)\ell+i-b$. For any $i \in [b+x':\tau-1]$, $h(y_i)$ is as shown below:
\begin{equation} \nonumber 
\begin{array}{cccccccccccccccccccc}
	0 &  & &  y_i & & & & i & & & &  \tau+y_i &   & & \tau+\delta \\
	(0 &\cdots& 0 & \alpha & 0 & \cdots & 0 & 1 & 0 & \cdots & 0 & X  & 0 & \cdots & 0)
	\end{array}
\end{equation}
%\bean
%\begin{array}{cccccccccccccccccccc}
%	& 0 & \multicolumn{2}{c}{\cdots} & (v_1-1)\ell+i-b &   \multicolumn{3}{c}{\cdots} & i & \multicolumn{3}{c}{\cdots} &  \tau+(v_1-1)\ell+i-b & \multicolumn{2}{c}{\cdots} & \tau+\delta &\\
%	( & 0 & \cdots & 0 & \alpha & 0 & \cdots & 0 & 1 &  0 & \cdots & 0 & X & 0 & \cdots & 0 &).
%\end{array}
%\eean
Let $S \subseteq [0:\tau+\delta]$ be the support of $h(t) \cap [b+x':\tau-1]$. We now look at the parity check equation given by $h(t) - \sum\limits_{i \in S} h(y_i)$. Clearly $y_i < t$ for all $i \in [b+x':\tau-1]$. It can be seen that the entries at indices $[b+x':\tau-1]$ of $h(t)$ are either $0$ or $1$. Hence $h(t) - \sum\limits_{i \in S} h(y_i)$ takes the following form:
%\bean
%\begin{array}{cccccccccccccccccccc}
%	& 0 & \multicolumn{2}{c}{\cdots} &  t &   \multicolumn{3}{c}{\cdots} & (\tau+(v_1-1)\ell+x_1) & \multicolumn{2}{c}{\cdots} &  \tau+t & \multicolumn{2}{c}{\cdots} & \tau+\delta &\\
%	( & X & \cdots & X & \alpha & 0 & \cdots & 0 & 1 &  X & \cdots & X & 0 & \cdots & 0 &).
%\end{array}.
%\eean
\begin{equation} \nonumber 
\begin{array}{ccccccccccccccccc}
	0 &  & &  t & &  & & \tau+(v-1)\ell+x' &   & & \tau+t & & & \tau+\delta \\
	(X &\cdots& X & \alpha & 0 & \cdots & 0 & 1 & X & \cdots & X & 0 & \cdots & 0)
	\end{array}
\end{equation}
Note that $\tau+(v-1)\ell+x' = \tau+t-\ell = t+b$. Thus there are $(b-1)$ zeros following index $t$ in the above equation. Therefore code symbol $c_t$ can be recovered by accessing code symbols with index $\le \tau+t$.

\subsection{Proof of Property R1}
Let $H^{(t)}$ be the parity check matrix of the punctured code obtained by deleting coordinates $[t+\tau+1:n-1]$ from the scalar code and let $\underline{h}_i^{(t)}$ denote the $i-$th column of $H^{(t)}$, for $i \in [0:\tau+t]$. To prove $R1$ property, it is enough to show that $\underline{h}_t^{(t)} \notin span\langle \{\underline{h}_i^{(t)} \mid i \in A \}\rangle$ for any $A \subset [t+1:\tau+t]$ with $|A|=a-1$, for all $t \in [0:\delta-1]$.
\subsubsection{$t=0$} In this case we need to look at $H^{(0)}$.
\bean
H^{(0)}=\left[ \begin{array}{c|ccccc}
	\alpha & X & \cdots   & \cdots & X & \alpha  \\
	\hline 
	0 & \multicolumn{2}{c}{ \multirow{3}{*}{$I_{(a-1)}$}} & \multicolumn{3}{c}{ \multirow{3}{*}{$\underbrace{\mathbf{C}}_{((a-1) \times (\tau+1-a))}$}}  \\
	\vdots  &  &   &   &  \\
	0 & & &  &  & 
\end{array}\right]. \\
\eean
Note that the last $(a-1)$ rows of $\underline{h}_0^{(0)}$ is all-zero. We also note that $H^{(0)}([1:a-1],[1:\tau])$ is the parity check matrix of an $[\tau,\tau+1-a]$ MDS code. Hence, it is not possible for any other $(a-1)$ columns of $H^{(0)}$ to linearly combine to obtain these $(a-1)$ zero entires, thus proving recoverability of $c_0$.
\subsubsection{$t \in [1:\delta-1]$} Here look at the $(a \times (\tau+1))$ matrix $\hat {H}_t=H(\{t\}\cup[\delta+1:b-1],[t:\tau+t])$ which is a sub-matrix of $H^{(t)}$.  Consider any $A \subseteq [1:\tau]$ with $|A|=a-1$. To prove $R1$ property it is sufficient to show that $0-$th column of $\hat {H}_t$ doesn't lie in the linear span of columns of $\hat {H}_t$ indexed by $A$.
\bean
\hat{H}_t=\left[ \begin{array}{c ccccc}
	\alpha & X & \cdots   & \cdots & X & 1  \\
	\hline 
	& \multicolumn{3}{c}{ \multirow{3}{*}{$H([\delta+1:b-1],[t:\tau])$}} & \multicolumn{2}{c}{ \multirow{3}{*}{$\underbrace{\mathbf{0}}_{((a-1) \times t)}$}}  \\
	&  &   &   &  \\
	& & &  &  & 
\end{array}\right], \\
\eean
where $X$ is either $0$ or $1$. If $A \cap [\tau-t+1:\tau]  \neq \phi$, then by MDS property the last $(a-1)$ entires of $0-$th column of $\hat {H}_t$ can not be obtained by linear combination of  $\{\hat {H}_t([1:a-1],j) \mid j \in A\}$. Now consider $A \subseteq [1:\tau-t]$. Suppose $\hat {H}_t([0:a-1],0)=\sum_{j \in A} \beta_j \hat {H}_t([0:a-1],j)$ where $\beta_j \in \mathbb{F}_{q^2}$. It can be argued using MDS parity property of $C$ matrix that there exists a unique linear combination of $\{\hat {H}_t([1:a-1],j) \mid j \in A\}$ with all coefficients in $\mathbb{F}_q \setminus \{0\}$ that result in $\hat {H}_t([1:a-1],0)$. Hence, $\beta_j \in \mathbb{F}_q \setminus \{0\}$, for all $j \in A$. Note that $\hat {H}_t(0,j) \in \mathbb{F}_q$ for all $j \in [1:\tau]$, whereas $\hat {H}_t(0,0)=\alpha \in \mathbb{F}_{q^2} \setminus \mathbb{F}_q$. So, $\hat {H}_t(0,0) \neq \sum_{j \in A} \beta_j \hat {H}_t(0,j)$, which is a contradiction. Thus, $0-$th column of $\hat {H}_t$ doesn't lie in span of columns in $A$.

\subsection{Proof of Property B2 }
It is clear from the definition of the B2 property that in order to prove B2 property it suffices to show  that the sub-matrix $H([0:b-1],[t:t+b-1])$ is invertible for any $t \in [\delta:\tau+1-a]$.  Before describing the general proof for invertibility of $H([0:b-1],[t:t+b-1])$, we first present some examples which illustrate our arguments. For $(a=2,b=5,\tau=12)$, the sub-matrix $H([0:4],[3:7])$ is as shown below: 
\bean
\left[ \begin{array}{cc|ccc}
	0 & 0 & 1 & 0 & 0   \\
	0 & 0 & 0 & 1 & 0   \\
	0 & 0 & 0 & 0 & 1  \\
	\hline 
	c_{0,1} & c_{0,2} & c_{0,3} & c_{0,4} & c_{0, 5}  \\
	c_{1,1} & c_{1,2} & c_{1,3} & c_{1,4} & c_{0,5}
\end{array}\right], 
\eean
where $c_{i,j}=H(\delta+i,a+j)$. This matrix is invertible since $\left[ \begin{array}{cc}
c_{0,1} & c_{0,2}  \\
c_{1,1} & c_{1,2} 
\end{array}\right]$
is a $(2 \times 2)$ sub-matrix of the parity check matrix of a $[13,11]$ MDS code. Now consider $H([0:4], [10:14])$ which has the following structure:
\bean
\left[ \begin{array}{ccc|cc}
	1 & 0 & \alpha & ~0~  & ~0~ \\
	\hline 
	0 & 1 & 0 & ~1~ & ~0~  \\
	1 & 0 & 0 & ~0~ & ~1~ \\
	\hline 
	c_{0,8} & c_{0,9} & c_{0,10} & ~0~ & ~0~   \\
	c_{1,8} & c_{1,9} & c_{1,10} & ~0~ & ~0~
\end{array}\right].
\eean
The above matrix is non-singular if  $M=\left[ \begin{array}{ccc}
1 & 0 & \alpha   \\
c_{0,8} & c_{0,9} & c_{0,10}   \\
c_{1,8} & c_{1,9} & c_{1,10}
\end{array}\right]$
has non-zero determinant. Clearly, $|M|= \begin{vmatrix}
c_{0,9} & c_{0,10}   \\
c_{1,9} & c_{1,10}
\end{vmatrix} + \alpha \begin{vmatrix}
c_{0,8} & c_{0,9}   \\
c_{1,8} & c_{1,9}
\end{vmatrix}$.
Since $\begin{vmatrix}
c_{0,9} & c_{0,10}   \\
c_{1,9} & c_{1,10}
\end{vmatrix} \in \mathbb{F}_q$, $\begin{vmatrix}
c_{0,8} & c_{0,9}   \\
c_{1,8} & c_{1,9}
\end{vmatrix} \ne 0$ and $\alpha \in \mathbb{F}_{q^2} \setminus \mathbb{F}_{q}$, we have $|M| \ne 0$. 

Now we move to the general proof of invertibility of  $H([0:b-1],[t:t+b-1])$ for any $t \in [\delta:\tau+1-a]$. Since $H([\delta:b-1],[0:\tau])$ is the parity-check matrix of a $[\tau+1,\tau+1-a]$ MDS Code, any $(a \times a)$ sub-matrix of it is invertible. We will use this fact repeatedly in the proof given below.
Let 
\bean 
\tau = v b + \ell ~~~\text{where}~~~ 0 \le \ell <b.
\eean 
Then $\tau+1-a=vb+\ell-a+1$. We divide the proof into multiple cases based on the range of value of $t$. In the below proof, we often ignore the sign of the determinant as it is irrelevant to invertibility of a matrix.
\subsection*{Case i) $t \in [\delta:(v-1)b]$}
In this case $t+b-1 \le vb-1=\tau-\ell-1$. We note that
\bean 
H([0:\delta-1],[\delta:vb-1]) 
	= \left[ \begin{array}{c|c|c|c|c|c}
	\underbrace{\mathbf{0}}_{\delta \times a} & \multirow{2}{*}{$~I_{\delta}~$}& \underbrace{\mathbf{0}}_{\delta \times a} & \multirow{2}{*}{$\cdots$} & \multirow{2}{*}{$~I_{\delta}~$} & \underbrace{\mathbf{0}}_{
		\delta  \times a} \\
	\end{array} \right].
\eean
From the structure of $H([0:\delta-1],[\delta:vb-1])$ given above it can be observed that  $H([0:\delta-1],[t:t+b-1])$, for $t \in [\delta:(v-1)b]$, is composed of $a$ all-zero columns and $\delta$ columns from identity matrix. Let $t = xb+\theta$ where $\theta < b$. 

Suppose $\theta \le \delta-1$, then the sub-matrix $H([0:b-1],[t:t+b-1])$ is of the following form:
\bean
\left[ \begin{array}{c|c|c}
	\multirow{2}{*}{$\underbrace{\mathbf{0}}_{(\theta \times (\delta-\theta))}$} & \multirow{4}{*}{~~~$\underbrace{\mathbf{0}}_{(\delta \times a)}$~~~} & \multirow{2}{*}{$I_{\theta}$}\\ & & \\ \cline{1-1} \cline{3-3}
	\multirow{2}{*}{$I_{(\delta-\theta)}$} & & \multirow{2}{*}{$\underbrace{\mathbf{0}}_{((\delta-\theta) \times \theta)}$}\\ 
	& & \\\hline 
	\multicolumn{3}{c}{H([\delta:b-1],[t:t+b-1])}
\end{array} \right].
\eean
%\bean
%\begin{array}{cccc|c|ccc}
%&	t & \cdots  & (x-1)b+\delta-1 & \cdots & xb &  \cdots & t+b-1\\ \hline
%\multirow{6}{*}{${\Bigg[ }$}	&  &  &  &  \multirow{6}{*}{$\underbrace{\mathbf{0}}_{(\delta \times a)}$}& 1 &   & \\
%  &	  &  & &  &  &  \ddots & \\
%  &	  &  &  &  &  &   & 1\\
%  & 	1 &  &  &  &  &   & \\
%  & 	  & \ddots &  &  &  &   & \\
%  & 	  &  & 1 &  &  &   & \\ \cline{2-8}
%&	  \multicolumn{7}{c}{G([0:a-1],[t:t+b-1])}
%\end{array}
%\eean
The determinant of the above matrix is equal to the determinant of $H([\delta:b-1],[xb+\delta:(x+1)b-1])$ which is an $(a\times a)$ sub-matrix of $H([\delta:b-1],[0:\tau])$ and is hence non-zero. 

For the case when $\theta \ge \delta$, the sub-matrix $H([0:b-1],[t:t+b-1])$  looks as shown below:
\bean
\left[ \begin{array}{c|c|c}
	\multirow{2}{*}{$\underbrace{\mathbf{0}}_{(\delta \times (b-\theta))}$} &   	\multirow{2}{*}{$~~I_{\delta}~~$} & 	\multirow{2}{*}{$\underbrace{\mathbf{0}}_{(\delta \times (\theta-\delta))}$}\\
	& & \\ \hline
	\multicolumn{3}{c}{H([\delta:b-1],[t:t+b-1])}
\end{array} \right].
\eean
%
%\bean
%\begin{array}{cc|ccc|c}
%	& t \ \cdots \ xb-1 & xb &  \cdots & xb+\delta-1 & xb+\delta  \cdots t+b-1\\ \cline{2-6}
%	\multirow{6}{*}{${\big[ }$}	& \multirow{6}{*}{$\underbrace{\mathbf{0}}_{(\delta \times (xb-t))}$ }& 1 &   & & \multirow{6}{*}{$\underbrace{\mathbf{0}}_{(\delta \times (a-xb+t))}$}\\
%	&	  &  &  \ddots & \\
%	&	  &  &   & 1 \\
%	& & & & &\\  \cline{2-6} 
%	&	  \multicolumn{5}{c}{G([0:a-1],[t:t+b-1])}
%\end{array}
%\eean
Let $A = [t:(x+1)b-1] \cup [(x+1)b+\delta:t+b-1]$.  The determinant of the sub-matrix shown above is equal to the determinant of the matrix $H([\delta:b-1], A)$. This determinant is non-zero since $H([\delta:b-1], A)$ is an $(a \times a)$ sub-matrix of parity check matrix of a $[\tau+1, \tau+1-a]$ MDS code. Thus we have completed the proof of invertibility of $H([0:b-1],[t:t+b-1])$ for all $t \in [\delta:(v-1)b]$.  
\subsection*{Case ii) $t \in [(v-1)b+1:(v-1)b+\ell]$}
Let $t = (v-1)b+\theta$. Therefore $1 \le \theta \le \ell$ and $t+b-1 = vb+\theta-1 \le vb+\ell-1$. Note that
\bean 
H([0:\delta-1],[(v-1)b:(vb+\ell-1)]) = \left[ \begin{array}{c|c|c}
	I_{\delta} & \underbrace{\mathbf{0}}_{
		\delta  \times a} & \mathbf{P}_{\delta, \ell}^a 
	\end{array} \right].
\eean
In this case the first $\theta$ columns of $\mathbf{P}_{\delta, \ell}^a$ are part of $H([0:\delta-1],[t:t+b-1])$. We first consider the case when $\theta \le \delta-1$. By definition of $\mathbf{P}_{\delta, \ell}^a$, we have $\mathbf{P}_{\delta, \ell}^a([0:\theta-1][0:\theta-1])=I_{\theta}$ for any $\theta \le \min\{\ell,\delta-1\}$. Hence the sub-matrix $H([0:b-1],[t:t+b-1])$ for the present case is as shown below:
\bean
\left[ \begin{array}{c|c|c}
	\multirow{2}{*}{~$\underbrace{\mathbf{0}}_{(\theta \times (\delta-\theta))}$~} & \multirow{4}{*}{~~~$\underbrace{\mathbf{0}}_{(\delta \times a)}$~~~} & \multirow{2}{*}{$I_{\theta}$}\\ & & \\ \cline{1-1} \cline{3-3}
	\multirow{2}{*}{$I_{(\delta-\theta)}$} & & \multirow{2}{*}{$\mathbf{P}_{\delta, \ell}^a([\theta:\delta-1],[0:\theta-1])$}\\ 
	& & \\\hline 
	\multicolumn{3}{c}{H([\delta:b-1],[t:t+b-1])}
\end{array} \right].
\eean
%\bean
%\begin{array}{cccc|c|ccc}
%	&	t & \cdots  & (v-1)b+\delta-1 & \cdots & vb &  \cdots & t+b-1\\ \hline
%	\multirow{6}{*}{${\Bigg[ }$}	&  &  &  &  \multirow{6}{*}{$\underbrace{\mathbf{0}}_{(\delta \times a)}$}& 1 &   & \\
%	&	  &  & &  &  &  \ddots & \\
%	&	  &  &  &  &  &   & 1\\
%	& 	1 &  &  &  & X &   \cdots & X\\
%	& 	  & \ddots &  &  &  \vdots&  \ddots & \vdots \\
%	& 	  &  & 1 &  & X &   \cdots& X\\ \cline{2-8}
%	&	  \multicolumn{7}{c}{G([0:a-1],[t:t+b-1])}
%\end{array}
%\eean
The determinant of this matrix is equal to the determinant of $H([\delta:b-1],[vb+\delta:vb-1])$ which is an $(a \times a)$ sub-matrix of the parity check matrix of an $[\tau+1,\tau+1-a]$ MDS code. Hence $H([0:b-1],[t:t+b-1])$ is non-singular. 

Now suppose $\ell \ge \theta \ge \delta$. Then, 
\bean 
\mathbf{P}_{\delta, \ell}^a([0:\delta-1],[0:\theta-1])=\Big[I_{\delta}~\underbrace{\mathbf{0}}_{(\delta \times (\theta-\delta))}\Big].
\eean 
Therefore the sub-matrix $H([0:b-1],[t:t+b-1])$ for this case has the following form:
\bean
\left[ \begin{array}{c|c|c}
	\multirow{2}{*}{$\underbrace{\mathbf{0}}_{(\delta \times (b-\theta))}$} &   	\multirow{2}{*}{$~~I_{\delta}~~$} & 	\multirow{2}{*}{$\underbrace{\mathbf{0}}_{(\delta \times (\theta-\delta))}$}\\
	& & \\ \hline
	\multicolumn{3}{c}{H([\delta:b-1],[t:t+b-1])}
\end{array} \right].
\eean
%\bean
%\begin{array}{cc|ccc|c}
%	& t \ \cdots \ vb-1 & vb &  \cdots & vb+\delta-1 & vb+\delta  \cdots t+b-1\\ \cline{2-6}
%	\multirow{6}{*}{${\big[ }$}	& \multirow{6}{*}{$\underbrace{\mathbf{0}}_{(\delta \times (xb-t))}$ }& 1 &   & & \multirow{6}{*}{$\underbrace{\mathbf{0}}_{(\delta \times (a-xb+t))}$}\\
%	&	  &  &  \ddots & \\
%	&	  &  &   & 1 \\
%	& & & & &\\  \cline{2-6} 
%	&	  \multicolumn{5}{c}{G([0:a-1],[t:t+b-1])}
%\end{array}
%\eean
Let $A = [t:vb-1] \cup [vb+\delta:t+b-1]$. The determinant of $H([0:b-1], [t:t+b-1])$ is equal to determinant of $H([\delta:b-1], A)$, which is non-zero since $A$ is an $a-$ element subset of $[0,\tau]$.
Thus we have argued that $H([0:b-1],[t:t+b-1])$ is non-singular for all $t \in [(v-1)b+1:(v-1)b+\ell]$.  

Before proving the next case, we present a property of  $P^a_{\delta,\ell}$ which will be helpful in the proof.
\blem
\label{P_property}
The sub-matrix formed by any $\ell$ consecutive rows of $P^a_{\delta,\ell}$ is invertible if $\ell \le \delta$.  
\elem 
\bprf %See full version of the paper at [].
Let $\delta = x \ell + u$ where $0 \le u < \ell$ and $x \ge 1$. Then $\mathbf{P}_{\delta, \ell}^a$ is of the following form:
\bean
\left[\begin{array}{cc}
	\multicolumn{2}{c}{I_{\ell}}\\
	\multicolumn{2}{c}{\vdots}\\
	\multicolumn{2}{c}{I_{\ell}}\\
	I_{u} & P_{u, \ell}^a([0:u-1],[u:\ell-1])
\end{array}\right].
\eean
It is easy to verify that any $\ell$ consecutive rows are linearly independent in the above matrix.
\eprf  
\subsection*{Case iii) $t \in [(v-1)b+\ell+1:(v-1)b+\delta]$}
This case is possible only when $\ell < \delta$. Let $t=(v-1)b+\ell+\theta$. Hence $1 \le \theta \le \delta-\ell-1$ and $t+b-1=vb+\ell+\theta-1=\tau+\theta-1$. Here the entire $\mathbf{P}_{\delta, \ell}^a$ is part of $H([0:\delta-1],[t:t+b-1])$.

The sub-matrix $H([0:b-1],[t:t+b-1])$ has the form:
\bean
\left[ \begin{array}{c|c|c|c}
	\mathbf{0} & \multirow{3}{*}{$\underbrace{\mathbf{0}}_{(\delta \times a)}$} & \multirow{3}{*}{$\mathbf{P}_{\delta, \ell}^a$} & B\\ \cline{1-1} \cline{4-4}
	\mathbf{0} & & & \mathbf{0}\\ \cline{1-1} \cline{4-4}
	I_{(\delta-\ell-\theta)} & & & \mathbf{0} \\ \hline
	\multicolumn{4}{c}{H([\delta:b-1],[t:t+b-1])}
	\end{array} \right],  
	\text{ where } B = \underbrace{\left[ \begin{array}{cccc}
		\alpha & & & \\
		& 1 & &\\
		& & \ddots & \\
		& & & 1
		\end{array} \right].}_{(\theta \times \theta)}
\eean
Note that $H([\delta:b-1],[(v-1)b+\delta:vb-1])$ is an $(a \times a)$ sub-matrix of $H([\delta:b-1],[0:\tau])$ and hence non-singular. Since $B$ is invertible, it  follows that $H([0:b-1],[t:t+b-1])$ is invertible if  $P_{\delta, \ell}^a ([\theta:\theta+\ell-1],[0:\ell-1])$ is invertible. Since  $\ell \le \delta-1$, it follows from Lemma~\ref{P_property} that $P_{\delta, \ell}^a ([\theta:\theta+\ell-1],[0:\ell-1])$ is invertible, thereby completing the proof for  $t \in [(v-1)b+\ell+1:(v-1)b+\delta]$.

%For the case $\theta=0$, the sub-matrix $H([0:b-1],[t:t+b-1])$ looks like:
%\bean
%\left[ \begin{array}{c|c|c}
%	\mathbf{0} & \multirow{2}{*}{$\underbrace{\mathbf{0}}_{(\delta \times a)}$} & \multirow{2}{*}{$\mathbf{P}_{\delta, \ell}^a$} \\ \cline{1-1} 
%	I_{(\delta-\ell)} &  \\  \hline
%	\multicolumn{3}{c}{H([\delta:b-1],[t:t+b-1])}
%\end{array} \right],
%\eean 
%This matrix is non-singular since  $H([\delta:b-1],[(v-1)b+\delta:vb-1])$ and $P_{\delta, \ell}^a ([0:\ell-1],[0:\ell-1])$ are non-singular by MDS property and Lemma~\ref{P_property} respectively. 
%We skip the regime of $t \in [(v-1)b+\delta:vb+\ell-a-1]$ and show the proof for remaining cases. We refer readers to full version of this paper at \cite{Small} for the complete proof. % that $H([0:b-1],[t:t+b-1])$ is invertible for the remaining range of $t$ values. 
%The skipped cases also follow arguments similar to the ones used above, but are more technical in nature.
\subsection*{Case iv) $t \in \{vb+\ell-a,vb+\ell-a+1\}$} 
If $t=vb+\ell-a$, then $t+b-1=\tau+\delta-1$ and  
the sub-matrix $H([0:b-1],[t:t+b-1])$ has the following structure:
\bean
\left[ \begin{array}{c|c|c}
	\multirow{4}{*}{ $H([0:\delta-1],[t:\tau-1])$} & \alpha & 0 \ \  \cdots \ \  0 \\ \cline{2-3}
	& 0  & \multirow{3}{*}{~~~~~$I_{\delta-1}$~~~~~}\\ 
	& \vdots &  \\ 
	& 0 &  \\ \hline
	\multicolumn{2}{c|}{\multirow{2}{*}{ $H([\delta:b-1],[t:\tau])$}} & \underbrace{\mathbf{0}}_{(a \times (\delta-1))} 
\end{array} \right].\\
\eean
The determinant of $H([0:b-1],[t:t+b-1])$ can be expanded along $0$-th row as $\pm \alpha*|H([\delta:b-1],[\tau-a:\tau-1])|+z$, where $z \in \mathbb{F}_q$. Note that $H([\delta:b-1],[\tau-a:\tau-1])$ has non-zero determinant as it is an $(a \times a)$ sub-matrix of MDS parity check matrix. Now since $\alpha \in \mathbb{F}_{q^2} \setminus \mathbb{F}_q$, we have $|H([0:b-1],[t:t+b-1])| \ne 0$. 

Now consider $t=vb+\ell-a+1$. In this case $t+b-1=\tau+\delta$ and the sub-matrix $H([\delta:b-1],[t:t+b-1])$ looks like: 
\bean
\left[ \begin{array}{c|c|c}
	\multirow{5}{*}{$H([0:\delta],[t:\tau-1])$} & \alpha & 0 \ \  \cdots \ \  0 \\ \cline{2-3}
	& 0 & \multirow{4}{*}{~~~~~$I_{\delta}$~~~~~}\\ 
	& \vdots & \\ 
	& 0 & \\ \cline{2-2}
	& H(\delta,\tau) &  \\ \hline
	\multicolumn{2}{c|}{\multirow{2}{*}{ $H([\delta+1:b-1],[t:\tau])$}} & \underbrace{\mathbf{0}}_{((a-1) \times \delta)} 
\end{array} \right].
\eean
The determinant of $H([0:b-1],[t:t+b-1])$ can be written as $\pm \alpha*|H([\delta+1:b-1],[\tau-a+1:\tau-1])|+z$, where $z \in \mathbb{F}_q$. Due to the MDS property, the $(a-1) \times (a-1)$ sub-matrix $H([\delta+1:b-1],[\tau-a+1:\tau-1])$ is invertible. Hence $|H([0:b-1],[t:t+b-1])| \ne 0$ as $\alpha \in \mathbb{F}_{q^2} \setminus \mathbb{F}_q$. 

\subsection*{Case v) $t \in [(v-1)b+\delta+1:vb]$} 
\subsubsection{$\ell<\delta$}Let $t = (v-1)b+\ell+\theta+1$. Then $\delta-\ell \le \theta \le b-\ell-1$ and $t+b-1=vb+\ell+\theta=\tau+\theta$. Since $n=\tau+\delta$ only $\theta \le \delta$ is possible. The cases $\theta=\delta$ and $\theta=\delta-1$ are already covered in case iv). Hence only $\theta < \delta-1$ is to be considered here. We also have $0 \le b-\theta-\ell-1 \le b-\delta-1<a$. The sub-matrix $H([0:b-1],[t:t+b-1])$ is thus of the form:
\bean
\left[ \begin{array}{c|c|c|c}
	\multirow{5}{*}{$\underbrace{\mathbf{0}}_{(\delta \times (b-\theta-\ell-1))}$} & \multirow{5}{*}{~~~$\mathbf{P}_{\delta, \ell}^a$~~~} & \alpha &  0 \ \cdots \cdots \ 0 \\ \cline{3-4}
	& & 0 & \multirow{3}{*}{~~~$I_{\theta}$~~~}\\ 
	& & \vdots & \\  \cline{4-4}
	& & \vdots & \multirow{2}{*}{$\underbrace{\mathbf{0}}_{((\delta-\theta-1) \times \theta)}$}\\ 
	& & 0 &	\\ \hline
	\multicolumn{3}{c|}{\multirow{2}{*}{ $H([\delta:b-1],[t:\tau])$}} & \underbrace{\mathbf{0}}_{(a \times \theta)} 
\end{array} \right].
\eean
The determinant of this matrix is equal to the determinant of the $(b-\theta) \times (b-\theta)$ matrix $R$ given below:
\bean
R = \left[ \begin{array}{c|c|c}
	\multirow{4}{*}{$\underbrace{\mathbf{0}}_{((\delta-\theta) \times (b-\theta-1-\ell))}$} & \multirow{4}{*}{~~~$\mathbf{P}'$~~~} & \alpha \\ \cline{3-3}% \multicolumn{2}{c}{B}\\  \cline{3-4}
	& & 0 \\ 
	& & \vdots \\ 
	& & 0 \\ \hline
	\multicolumn{3}{c}{H([\delta:b-1],[t:\tau])}  
\end{array} \right],
\eean
where $\mathbf{P}'=\mathbf{P}_{\delta, \ell}^a(\{0\} \cup [\theta+1:\delta-1], [0:\ell-1])$. Let $y = \delta-\theta-1$ and $\delta = x \ell + u$ where $0 \le u < \ell$. Note that $(b-\theta-1-\ell) = (a-(\ell-y))$. 
\paragraph{$y \le u$} Notice that $\ell-u \le \ell-y \le a$ here and therefore the reduced  $(b-\theta) \times (b-\theta)$ matrix $R$ is of the form shown below:
\bean
\left[\begin{array}{c|c|c|c|c}
	\multirow{3}{*}{$\underbrace{\mathbf{0}}_{(y+1) \times (a-(\ell-y))}$} & \multicolumn{3}{c|}{1 \ \ 0 \ \  \cdots \cdots \cdots \cdots \cdots \cdots  \ \ 0}  & \alpha \\ \cline{2-5}
	& \multirow{3}{*}{$\underbrace{\mathbf{0}}_{y \times (u-y)}$} & \multirow{3}{*}{~~~$I_y$~~~} & \multirow{3}{*}{$\underbrace{\mathbf{0}}_{y \times (\ell-u)}$}  & 0 \\
	& & & & \vdots\\ 
	& & & & 0\\ \hline
	\multicolumn{5}{c}{H([\delta:b-1],[t:\tau])}  
\end{array}\right].
\eean
The determinant of $R$ is given by $|R| = \pm |H([\delta:b-1],A\setminus\{vb\})|  \pm \alpha |H([\delta:b-1],A\setminus\{vb+\ell\})|$,
where $A = [t:vb+u-y-1] \cup [vb+u:vb+\ell]$ is a set of $a+1$ columns. The determinant $|R|$ is clearly non zero as $|H([\delta:b-1],A\setminus\{vb\})|$ and $|H([\delta:b-1],A\setminus\{vb+\ell\})|$ are both non zero in $\mathbb{F}_q$ and $\alpha \in \mathbb{F}_{q^2} \setminus \mathbb{F}_q$.
\paragraph{$y > u$} The reduced matrix $R$ has the following form:
\bean
\left[\begin{array}{c|c|c|c|c}
	\multirow{5}{*}{$\underbrace{\mathbf{0}}_{(y+1) \times (a-(\ell-y))}$} & \multicolumn{3}{c|}{1 \ \ 0 \ \  \cdots \cdots \cdots \cdots \cdots  \ \  0}  & \alpha \\ \cline{2-5}
	& \multirow{2}{*}{$\underbrace{\mathbf{0}}_{(y-u) \times u}$} & \multirow{4}{*}{$\underbrace{\mathbf{0}}_{y \times (\ell-y)}$}  & \multirow{2}{*}{$I_{y-u}$} & 0 \\
	& & & & \vdots \\ \cline{2-2} \cline{4-4}
	& \multirow{2}{*}{$I_{u}$} &  & \multirow{2}{*}{M} & \vdots \\
	& & & & 0\\ \hline
	\multicolumn{5}{c}{H([\delta:b-1],[t:\tau])}  
\end{array}\right],
\eean
where $M=P^a_{\delta,\ell}([\delta-u:\delta-1],[\ell-y-u:\ell-1])$.
Let $A = [t:t+a-\ell+y-1] \cup [vb+u:vb+u+\ell-y-1]$ be a set of size $a$. It can be seen that the determinant of R takes the form $|R| = \pm \alpha |H([\delta:b-1],A)| + z$, where $z \in \mathbb{F}_q$. Now from $\alpha \in \mathbb{F}_{q^2} \setminus \mathbb{F}_q$ and invertibility of $H([\delta:b-1],A)$ it follows that $|R| \ne 0$. 
\subsubsection{$\ell \ge \delta$} The proof for $t \in [(v-1)b+\delta+1:(v-1)b+\ell]$ is part of case ii). Therefore here we need to only consider $t \in [(v-1)b+\ell+1:vb]$. Let  $t = (v-1)b+\ell+1+\theta$, where $0 \le \theta \le b-\ell-1$.  Then $t+b-1=vb+\ell+\theta=\tau+\theta$ and $0 \le b-\ell-\theta-1<a$. Here we consider only $\theta < \delta-1$ as the other possible cases are handled in case iv). The sub-matrix $H([0:b-1],[t:t+b-1])$ is of the form:
\bean
\left[ \begin{array}{c|c|c|c|c}
	\multirow{5}{*}{$\underbrace{\mathbf{0}}_{(\delta \times (b-\theta-\ell-1))}$} & \multirow{5}{*}{~~~$\mathbf{I}_{\delta}$~~~} & 	\multirow{5}{*}{$\underbrace{\mathbf{0}}_{(\delta \times (\ell-\delta))}$} & \alpha & 0 \ \cdots \cdots \ 0 \\ \cline{4-5}% \multicolumn{2}{c}{B}\\  \cline{3-4}
	& & & 0 & \multirow{3}{*}{~~~$I_{\theta}$~~~}\\ 
	& & & \vdots &\\  \cline{5-5}
	& & & \vdots & \multirow{2}{*}{$\underbrace{\mathbf{0}}_{((\delta -\theta -1) \times \theta) }$}\\   
	& & & 0 &  \\ \hline
	\multicolumn{4}{c|}{\multirow{2}{*}{ $H([\delta:b-1],[t:\tau])$}} & \underbrace{\mathbf{0}}_{(a \times \theta)} 
\end{array} \right].
\eean
Since $\alpha \in \mathbb{F}_{q^2} \setminus \mathbb{F}_q$ it can be argued that determinant of $H([0:b-1],[t:t+b-1])$ is non-zero if the determinant of $(b-\theta-1) \times (b-\theta-1)$ matrix 
\bean
\left[ \begin{array}{c|c|c|c}
	\underbrace{\mathbf{0}}_{(\delta-\theta-1) \times (b-\ell-\theta-1)}  & \underbrace{\mathbf{0}}_{((\delta-\theta-1) \times (\theta+1))}  
	 &  \multirow{2}{*}{$I_{\delta-\theta-1}$} & \underbrace{\mathbf{0}}_{(\delta-\theta-1) \times (\ell -\delta)} \\
	& & & \\ \hline
	\multicolumn{4}{c}{H([\delta:b-1],[t:\tau-1])}
\end{array} \right]
\eean 
is non-zero.
The determinant of this matrix is equal to $|H([\delta:b-1],A)|$ where $A=[t:t+(b-\ell)-1] \cup [vb+\delta:vb+\ell-1]$ is a set of size $a$. Therefore the determinant is non-zero.

\subsection*{Case vi) $t \in [vb+1:vb+\ell-a-1]$} 
This case is possible only if $\ell \ge a+2$. Let $t+b-1=\tau+\theta = vb+\ell+\theta$, then $t = (v-1)b+
\ell+\theta+1$ and $ b-\ell \le \theta \le \delta-2$. The sub-matrix $H([0:b-1],[t:t+b-1])$ is of the form:
\bean
\left[\begin{array}{c|c|c}
	\multirow{5}{*}{$P_{\delta, \ell}^{a}([0:\delta-1],[\ell+\theta+1-b:\ell-1])$}& \alpha &  0 \ \cdots \cdots \ 0\\ \cline{2-3}
	& 0 & \multirow{3}{*}{$I_{\theta}$}\\ 
	& \vdots & \\\cline{3-3}
	& \vdots & \multirow{2}{*}{$\underbrace{\mathbf{0}}_{((\delta-\theta-1) \times \theta)}$}\\ 
	& 0 &  \\ \hline
	\multicolumn{2}{c|}{\multirow{2}{*}{ $H([\delta:b-1],[t:\tau])$}} & \underbrace{\mathbf{0}}_{(a \times \theta)} 
\end{array} \right].
\eean
As $\ell+\theta+1-b \ge 1$, the first row of above matrix has zeros in first $(b-\theta-1)$ columns. Therefore the matrix is invertible as long as the $(b-\theta-1) \times (b-\theta-1)$ reduced matrix $R$ shown below is invertible.
\bea
\label{eq:rmatrix}
R = \left[\begin{array}{c}
	P_{\delta, \ell}^a([\theta+1:\delta-1],[\ell+\theta+1-b:\ell-1])\\ \hline
	H([\delta:b-1],[t:\tau-1])
\end{array} \right]
\eea
\setcounter{subsubsection}{0}
\subsubsection{$\ell \ge \delta$} For this case $P_{\delta, \ell}^a = [I_{\delta} \ \mathbf{0}]$. Therefore matrix $R$ is of the form:
\bean
\left[ \begin{array}{c|c|c}
	\underbrace{\mathbf{0}}_{((\delta-\theta-1) \times (b-\ell))} & \multirow{2}{*}{~~$I_{\delta-\theta-1}$~~} & \underbrace{\mathbf{0}}_{((\delta-\theta-1) \times (\ell -\delta))} \\
	& & \\ \hline
	\multicolumn{3}{c}{H([\delta:b-1],[t:\tau-1])}
\end{array} \right].
\eean
The determinant of $R$ is equal to $|H([\delta:b-1],A)|$ where $A=[t:t+(b-\ell)-1] \cup [vb+\delta:vb+\ell-1]$ is set of $a$ columns. This is clearly non zero.
\subsubsection{$\ell < \delta$} Let $y = \delta-\theta-1$. Note that $\ell \ge b-\theta = \delta-\theta +a$ and hence we have $0<y<\ell$. We will first examine the structure of the $y \times (a+y)$ sub-matrix of $P_{\delta, \ell}^a$ corresponding to last $y$ rows and last $(a+y)$ columns that appears in the reduced matrix $R$. To do that we define the variables $v_0, v_1, v_2, \cdots$ where $v_0 = \delta$, $v_1 = \ell$ and
\bean
v_{i} = \begin{cases}
	v_{i-2} \mod v_{i-1} & i \text{ even},\\
	v_{i-2} \mod (v_{i-1} + a) & \text{otherwise}.
\end{cases}
\eean
If $v_{i} = 0$, we set $D = i$. For $i$ odd if $v_{i} \ge v_{i-1}$ also, we set $D = i$ and $v_D = 0$. By this definition:
\bean
\delta = v_0 > \ell = v_1 > v_2 > v_3 \cdots > v_D = 0.
\eean
We note that $D \ge 2$ always.
\subsubsection*{An Example- $\delta=15, \ell=4, a=2$} Here $v_0=\delta=15$ and $v_1=\ell=4$. Now $v_2=v_0 \mod v_1= 15 \mod 4 =3$. Then $v_3= v_1 \mod (v_2+a)= 4 \mod (3+2)=4$. This means $v_3>v_2$, therefore $D=3$ and $v_3=0$. Thus for this example case $v_0=15 > v_1=4 > v_2=3 > v_3=0$. \\ 

The indices $v_0, \cdots, v_D$ help describe the structure of $P_{\delta, \ell}^a$. 
For $i$ even,
\bean
P_{v_{i-2},v_{i-1}}^a &=& \begin{cases} \left[\begin{array}{c}
		I_{v_{i-1}}\\
		\vdots\\
		I_{v_{i-1}}\\
		P_{v_{i},v_{i-1}}^a
	\end{array}\right] &  i \ne D\\ \ \\
	\left[\begin{array}{c}
		I_{v_{i-1}}\\
		\vdots\\
		I_{v_{i-1}}\\
	\end{array}\right] &  i = D\\
\end{cases} 
\eean 
For $i$ odd and $i \ne D$ ,
\bean 
P_{v_{i-1}, v_{i-2}}^a=\left[I_{v_{i-1}} \underbrace{\mathbf{0}}_{v_{i-1} \times a} \cdots I_{v_{i-1}} \underbrace{\mathbf{0}}_{v_{i-1} \times a} P_{v_{i-1},v_{i}} \right]. 
\eean 
For odd $D$,  
\bean 
P_{v_{D-1}, v_{D-2}}^a=\left[I_{v_{D-1}} ~\underbrace{\mathbf{0}}_{v_{D-1} \times a}~ \cdots~ I_{v_{D-1}} ~\underbrace{\mathbf{0}}_{v_{D-1} \times \mu} \right]
\eean 
where 
\bean 
\mu=\begin{cases}
	a ~~~\text{if}~~~v_{D-2} \mod (v_{D-1}+a)=0, \\
	v_{D-2} \mod (v_{D-1}+a)-v_{D-1}~~~\text{otherwise}.
\end{cases}.
\eean   \\ 
Look at $y \times (a+y)$ sub-matrix of $P_{\delta, \ell}^a$ appearing in $R$. Since $0 < y<\ell$, there exists some $i \in [2:D]$ such that $v_{i-1} > y \ge v_{i}$. We prove the $i$ even and odd cases separately.
\paragraph{$i$ even} For even $i \ge 4$, note that $v_{i-2} > v_{i-1} > y$ and $v_{i-3} \ge v_{i-1}+v_{i-2}+a > a+y$. Hence, the $y \times (a+y)$ sub-matrix of $P_{\delta,\ell}^a$ under consideration  is  $y \times (a+y)$  sub-matrix of $P_{v_{i-2},v_{i-3}}^a$ corresponding to last $(a+y)$ columns and last $y$ rows. For even $i \ge4$,
\bean
P_{v_{i-2},v_{i-3}}^a = \left[\begin{array}{cccccccc}
	\multirow{2}{*}{$I_{v_{i-2}}$} & \underbrace{\mathbf{0}}_{v_{i-2} \times a} & \multirow{2}{*}{$\cdots$} & 	\multirow{2}{*}{$I_{v_{i-2}}$} & \underbrace{\mathbf{0}}_{v_{i-2} \times a} & \multirow{2}{*}{$P_{v_{i-2},v_{i-1}}^a$}
\end{array}\right].
\eean
For the case when even $i=D\ge4$, this matrix looks as:
\bean
P_{v_{D-2}, v_{D-3}}^a = \left[\begin{array}{ccccc|c}
	\multirow{4}{*}{$I_{v_{D-2}}$} & \multirow{4}{*}{$\underbrace{\mathbf{0}}_{v_{D-2} \times a}$} & \multirow{4}{*}{$\cdots$} & 	\multirow{4}{*}{$I_{v_{D-2}}$} & \multirow{4}{*}{$\underbrace{\mathbf{0}}_{v_{D-2} \times a}$} & I_{v_{D-1}}\\ \cline{6-6}
	& & & & & \vdots\\ \cline{6-6}
	& & & & & I_{v_{D-1}}\\ 
\end{array}\right].
\eean
If $i=D=2$, then 
\bean
P_{v_0, v_1}^a = \left[\begin{array}{c}
	I_{v_1} \\ 
	\vdots \\ 
	I_{v_1} 
\end{array}\right].
\eean
Note that when $i=D$ we have $V_{D-1}>y>0$. Hence, for even $i=D$ the reduced sub-matrix $R$ shown in equation \eqref{eq:rmatrix} has the following form:
\bean
R = \left[\begin{array}{c|c}
	\multirow{2}{*}{~~~$\underbrace{0}_{(y \times a)}$~~~} & \multirow{2}{*}{$I_{y}$}\\ 
	& \\ \hline
	\multicolumn{2}{c}{H([\delta:b-1],[t:t+a+y-1])}
\end{array}\right].
\eean
The determinant of $R$ is same as determinant of $H([\delta:b-1],[t:t+a-1])$. This is clearly non-zero by the definition of $H$. 

For the case when $i$ is even with $4 \le i < D$, the matrix $P_{v_{i-2},v_{i-3}}^a$ is of the following form:
\bean
P_{v_{i-2}, v_{i-3}}^a = \left[\begin{array}{ccccc|c}
	\multirow{4}{*}{$I_{v_{i-2}}$} & \multirow{4}{*}{$\underbrace{\mathbf{0}}_{v_{i-2} \times a}$} & \multirow{4}{*}{$\cdots$} & 	\multirow{4}{*}{$I_{v_{i-2}}$} & \multirow{4}{*}{$\underbrace{\mathbf{0}}_{v_{i-2} \times a}$} & I_{v_{i-1}}\\ \cline{6-6}
	& & & & & \vdots\\ \cline{6-6}
	& & & & & I_{v_{i-1}}\\ \cline{6-6}
	& & & & & P_{v_i, v_{i-1}}^a\\ 
\end{array}\right].
\eean
If $i$ is even and $v_{i-1} \le a+y$, then $i \ge 4$ since $v_1=\ell > a+y$. For the case when $i<D$ is even with $v_{i-1} \le a+y$, the $(y \times (a+y))$ sub-matrix of interest has the form:
\bean
\left[
\begin{array}{c|c|c}
	\multirow{3}{*}{$\underbrace{0}_{y \times (a+y-v_{i-1})}$} & \multirow{2}{*}{$\underbrace{\mathbf{0}}_{(y-v_i) \times (v_{i-1}-(y-v_i))}$} & \multirow{2}{*}{$I_{y-v_i}$}\\
	& & \\ \cline{2-3}
	& \multicolumn{2}{c}{P_{v_i, v_{i-1}}^a}
\end{array} \right].
\eean
In this case the reduced sub-matrix $R$ shown in equation \eqref{eq:rmatrix} is as shown below:
\bean
R = \left[\begin{array}{c|c|c}
	\multirow{3}{*}{$\underbrace{0}_{y \times (a+y-v_{i-1})}$} & \multirow{2}{*}{$\underbrace{\mathbf{0}}_{(y-v_i) \times (v_{i-1}-(y-v_i))}$} & \multirow{2}{*}{$I_{y-v_i}$}\\
	& & \\ \cline{2-3}
	& \multicolumn{2}{c}{P_{v_i, v_{i-1}}^a}\\ \hline
	\multicolumn{3}{c}{H([\delta:b-1],[t:t+a+y-1])}
\end{array}\right].
\eean
The determinant of $R$ is same as the determinant of the matrix shown below:
\bean
R' = \left[\begin{array}{c|c}
	\underbrace{0}_{v_i \times (a+y-v_{i-1})} & P_{v_i, v_{i-1}}^a([0:v_i-1],[0:v_{i-1}-y+v_i-1])\\ \hline
	\multicolumn{2}{c}{H([\delta:b-1],[t:t+a+v_i-1])}
\end{array}\right],
\eean 
Since $v_{i-1}-y \le a$ in this case,
\bean 
P_{v_i, v_{i-1}}^a([0:v_i-1],[0:v_i+(v_{i-1}-y-1)]) = \left[\begin{array}{cc}
	\multirow{2}{*}{$I_{v_i}$} & \underbrace{\mathbf{0}}_{v_i \times (v_{i-1}-y)} 
\end{array}\right].
\eean  
Therefore the determinant of $R$ is equal to $|H([\delta:b-1], A)|$ where $A = [t:t+a+y-v_{i-1}-1] \cup [t+a+y-v_{i-1}+v_i:t+a+v_i-1]$ is a set of $a$ columns. It is clear to see that it is hence non-zero. 

Now consider the case when $i<D$ is even with $v_{i-1} > a+y$. Then the $(y \times (a+y))$ sub-matrix of $P^a_{\delta,\ell}$ appearing in $R$ is of the following form:
\bean
\left[\begin{array}{c|c}
	\multirow{2}{*}{~~~$\underbrace{\mathbf{0}}_{(y-v_i) \times (a+v_i)}$~~~} & \multirow{2}{*}{$I_{y-v_i}$}\\
	& \\ \hline
	\multicolumn{2}{c}{P_{v_i, v_{i-1}}^a([0:v_i-1],[v_{i-1}-a-y:v_{i-1}-1])}
\end{array}\right].
\eean
In this case the reduced sub-matrix $R$ is of the form:
\bean
R = \left[\begin{array}{c|c}
	\multirow{2}{*}{~~~$\underbrace{\mathbf{0}}_{(y-v_i) \times (a+v_i)}$~~~} & \multirow{2}{*}{$I_{y-v_i}$}\\
	& \\ \hline
	\multicolumn{2}{c}{P_{v_i, v_{i-1}}^a([0:v_i-1],[v_{i-1}-a-y:v_{i-1}-1])} \\ \hline
	\multicolumn{2}{c}{H([\delta:b-1],[t:t+a+y-1])}
\end{array}\right].
\eean
The determinant of $R$ is same as the determinant of matrix $R'$ defined as:
\bean
\label{eq:Rmatrix3} R' = \left[\begin{array}{c}
	P_{v_i, v_{i-1}}^a([0:v_i-1],[v_{i-1}-a-y:v_{i-1}+v_i-y-1]) \\ \hline
	H([\delta:b-1],[t:t+a+v_i-1])
\end{array}\right].
\eean
It can be seen that $(a+v_i)$ columns of $P_{v_i, v_{i-1}}^a$ appear in matrix $R'$ and 
\bean 
P_{v_i, v_{i-1}}^a = \left[\begin{array}{cccccc}
	I_{v_i} & \underbrace{\mathbf{0}}_{(v_i \times a)} & \cdots & I_{v_i} &  \underbrace{\mathbf{0}}_{(v_i \times a)} & P_{v_i,v_{i+1}}^a
\end{array}\right].
\eean
\bit
\item If $y-v_i \ge v_{i+1}$ the $(v_i \times (a+v_i))$ sub-matrix of $P_{v_i, v_{i-1}}^a([0:v_i-1],[v_{i-1}-a-y:v_{i-1}+v_i-y-1])$ contained in $R'$ is a sub-matrix that is composed of $(a+v_i)$ consecutive columns from the matrix:
\bean
\left[ \begin{array}{ccccc}
	I_{v_i} & \underbrace{\mathbf{0}}_{(v_i \times a)} & \cdots & I_{v_i} &  \underbrace{\mathbf{0}}_{(v_i \times a)} \end{array} \right]
\eean
There will be $a$ all-zero columns among the $a+v_i$ consecutive columns of the above matrix appearing in $R'$. Hence the determinant of $R'$ is equal to determinant of sub-matrix composed of $a$ columns of $H([\delta:b-1],[t:t+a+v_i-1])$. Therefore $|R'| \ne 0$.

\item Otherwise ie., $y-v_i < v_{i+1}$, then the $a+v_i$ columns of $P_{v_i, y_{i-1}}^a$ that are part of $R'$ also include elements from $P_{v_i, v_{i+1}}^a$. Let $y_1 = v_{i+1}+v_i-y$. Then the matrix $R'$  is of the form shown below:
\bean
\left[\begin{array}{c|c|c}
	\mathbf{0} & \multirow{2}{*}{$\underbrace{\mathbf{0}}_{v_i \times a}$} & \multirow{2}{*}{$P_{v_i, v_{i+1}}^a([0:v_i-1],[0:y_1-1])$}\\ \cline{1-1}
	I_{v_i-y_1}& & \\ \hline
	\multicolumn{3}{c}{H([\delta:b-1],[t:t+a+v_i-1])}
\end{array} \right]
\eean
This matrix is invertible as $P_{v_i, v_{i+1}}^a([0:y_1-1],[0:y_1-1]) = I_{y_1}$ and any $a$ columns of $H([\delta:b-1],[t:t+a+v_i-1])$ are linearly independent.
\eit 
\paragraph{$i$ odd} Since $v_1=\ell>y$ we have $i \ge 3$ if $i$ is odd. For odd $i <D$, the $y \times (a+y)$ sub-matrix of $P_{\delta, \ell}^a$ appearing in $R$ is  $y \times (a+y)$  sub-matrix of $P_{v_{i-1},v_{i-2}}^a$ corresponding to last $(a+y)$ columns and last $y$ rows. This is because $v_{i-2} \ge v_{i-1}+a > a+y$ and $v_{i-1} > y$. For odd $i < D$. 
\bean
P_{v_{i-1}, v_{i-2}}^a  = \left[\begin{array}{cccccccc}
	\multirow{2}{*}{$I_{v_{i-1}}$} & \underbrace{\mathbf{0}}_{v_{i-1} \times a} & \multirow{2}{*}{$\cdots$} & 	\multirow{2}{*}{$I_{v_{i-1}}$} & \underbrace{\mathbf{0}}_{v_{i-1} \times a} & \multirow{2}{*}{$P_{v_{i-1},v_{i}}$}
\end{array}\right].
\eean
Therefore the sub-matrix $R$ appearing in equation \eqref{eq:rmatrix} can be written as:
\bean
\left[ \begin{array}{c|c|c}
	\mathbf{0} & \multirow{2}{*}{$\underbrace{\mathbf{0}}_{y \times a}$} & \multirow{2}{*}{$P_{v_{i-1}, v_i}^a ([v_{i-1}-y:v_{i-1}-1],[0:v_i-1])$}\\ \cline{1-1}
	I_{y- v_i} & & \\ \hline
	\multicolumn{3}{c}{H([\delta:b-1],[t:\tau-1])}
\end{array} \right].
\eean
The determinant of this matrix is equal to:
\bean
|H([\delta:b-1],[t+y-v_i:t+a+y-v_i])| * \\ |P_{v_{i-1},v_i}^a([v_{i-1}-y:v_{i-1}-y+v_i-1],[0:v_i-1])|.
\eean
The first term in this determinant is non-zero by MDS property. The second term corresponds to determinant of sub-matrix of $P_{v_{i-1},v_i}^a$ that is formed by picking $v_i$ consecutive rows. By the property of $P_{v_{i-1}, v_i}^a$ given in Lemma~\ref{P_property} this is non-zero.\\

For the case odd $i=D$ with $v_{D-2} \mod (v_{D-1}+a)=0$, we have $v_{D-1}>y$ and $v_{D-2} \ge v_{D-1}+a>a+y$ and hence the $y \times (a+y)$ sub-matrix appearing in $R$ is contained in  
\bean
P_{v_{D-1}, v_{D-2}}^a  = \left[\begin{array}{ccccccc}
	\multirow{2}{*}{$I_{v_{D-1}}$} & \underbrace{\mathbf{0}}_{v_{D-1} \times a} & \multirow{2}{*}{$\cdots$} & 	\multirow{2}{*}{$I_{v_{D-1}}$} & \underbrace{\mathbf{0}}_{v_{D-1} \times a}
\end{array}\right].
\eean
The reduced matrix $R$ defined in \eqref{eq:rmatrix} takes the form
\bean
R = \left[\begin{array}{c|c}
	~~~~~I_{y}~~~~~ & \underbrace{0}_{(y \times a)} \\ \hline
	\multicolumn{2}{c}{H([\delta:b-1],[t:t+a+y-1])}
\end{array}\right].
\eean
Hence, the determinant of $R$ is equal to determinant of $H([\delta:b-1],[t+y:t+a+y-1])$, which is clearly non-zero.

Now consider the case when odd $i=D$ with $v_{D-2} \mod (v_{D-1}+a) \ge v_{D-1}$. 
\bit
\item For the case when $v_{D-2} \ge a+y$, the $(y \times (a+y))$ sub-matrix we are interested in is a sub-matrix of $P_{v_{D-1},v_{D-2}}^a$ and 
\bean
P_{v_{D-1},v_{D-2}}^a = \left[\begin{array}{cccccc}
	\multirow{2}{*}{$I_{v_{D-1}}$} & \underbrace{\mathbf{0}}_{v_{D-1} \times a} & \multirow{2}{*}{$\cdots$} & 	\multirow{2}{*}{$I_{v_{D-1}}$} & \underbrace{\mathbf{0}}_{v_{D-1} \times p} 
\end{array}\right],
\eean
where $p = v_{D-2} \mod(a+v_{D-1}) - v_{D-1}$.
The $y \times (a+y)$ sub-matrix comprised of last $y$ rows, last $(a+y)$ columns is of the form shown below:
\bean
\left[\begin{array}{ccc}
	\underbrace{\mathbf{0}}_{y \times (a-p)}  & 	I_{y} & \underbrace{\mathbf{0}}_{y \times p} 
\end{array}\right]
\eean
Therefore the sub-matrix appearing in equation \eqref{eq:rmatrix} can be written as:
\bean
R = \left[ \begin{array}{c|c|c}
	\underbrace{\mathbf{0}}_{y \times (a-p)}  & 	I_{y} & \underbrace{\mathbf{0}}_{y \times p} \\ \hline
	\multicolumn{3}{c}{H([\delta:b-1],[t:\tau-1])}
\end{array} \right].
\eean
The determinant of this matrix is equal to $|H([\delta:b-1],A)|$ where $A = [t:t+a-p-1] \cup [\tau-p:\tau-1]$, which is non-zero since $A$ is of size $a$.
\item For the case when $v_{D-2} < a+y$,  $P_{v_{D-1},v_{D-2}}^a$ is as shown below:
\bean
P_{v_{D-1},v_{D-2}}^a = \left[\begin{array}{cccccc}
	\multirow{2}{*}{$I_{v_{D-1}}$} & \underbrace{\mathbf{0}}_{v_{D-1} \times p} 
\end{array}\right],
\eean
where  $p = v_{D-2} - v_{D-1}$. Since $v_1=\ell> a+y$, this case occurs only when $D \ge 5$. In this case, the $(y \times (a+y))$ sub-matrix of $P_{\delta, \ell}^a$ appearing in $R$ is the sub-matrix of $P_{v_{D-3},v_{D-4}}^a$  corresponding to last $(a+y)$ columns and last $y$ rows. 
\bean
P_{v_{D-3},v_{D-4}}^a  = \left[\begin{array}{ccccc|c}
	\multirow{4}{*}{$I_{v_{D-3}}$} & \multirow{4}{*}{$\underbrace{\mathbf{0}}_{v_{D-3} \times a}$} & \multirow{4}{*}{$\cdots$} & 	\multirow{4}{*}{$I_{v_{D-3}}$} & \multirow{4}{*}{$\underbrace{\mathbf{0}}_{v_{D-3} \times a}$} & I_{v_{D-2}}\\ \cline{6-6}
	& & & & & \vdots\\ \cline{6-6}
	& & & & & I_{v_{D-2}}\\ \cline{6-6}
	& & & & & P_{v_{D-1}, v_{D-2}}^a\\ 
\end{array}\right].
\eean
Thus the $y \times (a+y)$ sub-matrix of interest has the form shown below:
\bean
\left[\begin{array}{ccc}
	\underbrace{\mathbf{0}}_{y \times (a-p)}  & 	I_{y} & \underbrace{\mathbf{0}}_{y \times p} 
\end{array}\right]. 
\eean
Therefore the sub-matrix appearing in equation \eqref{eq:rmatrix} can be written as:
\bean
R = \left[ \begin{array}{c|c|c}
	\underbrace{\mathbf{0}}_{y \times (a-p)}  & 	I_{y} & \underbrace{\mathbf{0}}_{y \times p} \\ \hline
	\multicolumn{3}{c}{H([\delta:b-1],[t:\tau-1])}
\end{array} \right].
\eean
The determinant of this matrix is equal to $|H([\delta:b-1],A)|$ where $A = [t:t+a-p-1] \cup [\tau-p:\tau-1]$, which is non-zero.
\eit 

\subsection{Proof of R2 property}
If we prove that columns $\{H([0:b-1],j) \mid j \in A\}$ are linearly independent for any set $A \subseteq [\delta:\tau+\delta]$ with $|A|=a$, then R2 property follows. For $\delta=0$, the scalar code reduces to a $[\tau+1,\tau+1-a]$ MDS code and proof is straightforward. Hence we need to consider only $\delta>0$, for which have $0 \notin A$. If $A \cap [1:\tau] \ne \phi$, observe that columns of $H([\delta:b-1],A)$ are either all-zero columns or distinct columns from an MDS parity check matrix. Hence the $a$ columns  $\{H([\delta:b-1],j) \mid j \in A\}$ are not linearly dependent. If $A \cap [0:\tau]$ is empty, then $A \subseteq [\tau+1: \tau+\delta]$. By B2 property for $t=\tau+1-a$, it follows that the columns of $H([0:b-1],A)$ are linearly independent. This completes the proof of R2 property. 

\bibliographystyle{IEEEtran}
\bibliography{streaming}
\end{document}